\documentclass[oldversion]{aa} 
\usepackage{graphicx}
\usepackage{txfonts}

\usepackage{natbib}
\bibpunct{(}{)}{;}{a}{}{,} 

\begin{document}
   \title{X-ray variability time scales in Active Galactic Nuclei}

   \subtitle{}

   \author{W. Ishibashi
          \inst{1,2}
          \and
          T. J.-L. Courvoisier
          \inst{1,2}
          }

   \institute{ ISDC Data Centre for Astrophysics, ch. d'Ecogia 16, 1290 Versoix, Switzerland 
    \and Geneva Observatory, Geneva University, ch. des Maillettes 51, 1290 Sauverny, Switzerland \\
    e-mail: Wakiko.Ishibashi@unige.ch, Thierry.Courvoisier@unige.ch 
             }

   \date{Received; accepted}

 
  \abstract{X-ray variability in Active Galactic Nuclei (AGN) is commonly analysed in terms of the Power Spectral Density (PSD). The break observed in the power spectrum can be interpreted as a characteristic X-ray variability time scale. Here we study variability properties within the framework of clumpy accretion flows, in which shocks between accreting elements account for the UV and X-ray emissions. We derive a characteristic X-ray time scale, $\tau_{X}$, and compare it with the measured PSD break time scale, $T_{B}$. A quite good agreement is found in both magnitude and trend. In particular, the model dependence on black hole mass and accretion rate precisely reproduces the empirical relation obtained by McHardy et al. (2006). We suggest a possible physical interpretation of the break time scale and briefly discuss the related aspects of optical/UV variability and correlations between different wavelengths. }


   \authorrunning{ W. Ishibashi \and T. J.-L. Courvoisier}
   \titlerunning{X-ray variability timescales in AGN}

 \maketitle

\section{Introduction}

Variability in Active Galactic Nuclei (AGN) covers a wide range of time scales and amplitudes  over the entire electromagnetic spectrum. 
The observed light curves are characterized by aperiodic and featureless fluctuations,
suggesting that the variability mechanism is of random nature.

Early studies of X-ray variability focused on the search of characteristic time scales (\citet{McH_2001} and references therein). 
One of the methods used to study the temporal structure of the variations is the Power Spectral Density (PSD) analysis. 
PSD results are mainly derived from high quality X-ray light curves provided by the Rossi X-ray Timing Explorer (RXTE) and XMM-Newton.  
The observed power spectrum is generally modeled by a power law of the form $P_{\nu} \propto \nu^{\alpha}$, where $\nu = 1/T$ is the temporal frequency.
At high frequencies (short time scales), the PSD presents a steep slope of $\alpha$$\sim$-2, with no characteristic time scale.  
At low frequencies (long time scales), the PSD is characterized by a slope of $\alpha$$\sim$-1, representing flicker noise. A characteristic time scale can be defined, associated with this bend in the power spectrum.
Estimations of the PSD bend or break time scale are obtained by fitting broken power laws to the observed PSD. 

Subsequent studies showed that the characteristic break time scale $T_{B}$ is related to the black hole mass $M_{BH}$: the larger the central mass, the longer the characteristic time scale, and a linear scaling (of the form $T_{B} \propto M_{BH}$) has been proposed (\citet{M_et_2003}, \citet{P_2004}).
But a non negligible scatter has soon been noticed, with narrow-line objects not fitting into the general picture. 
Narrow-line Seyfert 1 galaxies (NLS 1) form a particular class of AGNs, characterized by rapid and large variability combined with other peculiar spectral properties. 
NLS 1 galaxies are usually believed to be powered by small black holes accreting at high rates, close to the Eddington limit. 
The fact that the break time scales are shorter in NLS 1 objects when compared with  classical broad-line galaxies of the same mass, suggests that the break time scale depends on a second parameter, such as accretion rate or black hole spin. (\citet{McH_et_2004}, \citet{U_McH_2005}). 
Following this hypothesis, \citet{McH_et_2006} proposed a sort of `fundamental plane' relating variability time scale, black hole mass, and accretion rate. In particular, they obtained an observational scaling relationship in which the break time scale explicitly depends on 
both black hole mass and accretion rate. 

The PSD break time scale is expected to provide insights into the emission mechanisms, and several physical interpretations have been proposed. 
Various accretion disc time scales have been compared with the measured PSD time scales, but with no conclusive result (\citet{M_et_2003}, \citet{P_2004}). 
The currently favored interpretation is based on the inner propagating fluctuation models (\citet{L_1997}, \citet{C_et_2001}). 
 
Cooling processes responsible for the emitted radiation are known in considerable detail, while the process of transfer of gravitational energy into radiative energy is still poorly understood.  
In standard accretion disc models, it is provided by some form of viscosity parametrized by the $\alpha$ parameter (\citet{S_S_1973}).
Another likely source for heating of the accreting gas is given by dissipative processes, such as shocks. In previous papers (\citet{C_T_2005}, \citet{I_C_2009}, hereafter Paper I), we discussed the cascades of shocks model in which radiation is emitted as a result of shocks between elements (clumps) forming the accretion flow.
In this picture, clumps move with velocities determined by the gravitational field of the central black hole following different orbits depending on the different initial conditions. 
In the central regions, colliding clumps are characterized by high relative velocities. 
When such high-velocity flows converge, shocks will result. This process releases the bulk of the clump kinetic energy converting it into radiation.
Optically thick and optically thin shocks then account for the observed optical/UV and X-ray emissions, respectively. 
Characteristic time scales can be derived within this framework.
Here we compare model time scales with observed variability time scales and suggest a possible interpretation of the break seen in the power spectrum. 

The present paper is organized as follows. 
Some elements of clumpy accretion flows are recalled in Sect. 2. 
We derive X-ray variability time scales and present model results in Sect. 3; comparison with observational results are given in Sect. 4.  
We then discuss some aspects of optical/UV variability in Sect. 5 and correlations between different wavelengths in Sect. 6. The main results are summarized in the Conclusion.


\section{Cascades of shocks in clumpy accretion flows}

In Paper I, we studied the properties of the inhomogeneous accretion flow formed by interacting clumps of matter, where shocks between elements and the subsequent evolution are at the origin of the radiation. 

A collision between two clumps of several solar masses, $M_{c} = M_{33} \cdot 10^{33}$g, moving at the local free-fall velocity at a typical distance of $\sim$$100 R_{S}$ (parametrized in units of the Schwarzschild radius $R= \zeta \cdot R_{S}$ where $\zeta = \zeta_{UV} \cdot 100$) leads to an optically thick shock. 
Following this optically thick shock the resulting gas cloud expands rapidly, with a fraction $\eta_{rad} = \eta_{1/3} \cdot \frac{1}{3}$ of the kinetic energy being radiated at the photosphere. 
The time it takes for the collision energy to be radiated is given by the expansion time
\begin{equation}
\tau_{exp} \cong 10^{6} M_{33}^{1/2} \zeta_{UV}^{1/4} \, \textrm{s} \, .
\end{equation}

We assume a spherical expansion with the photospheric radius given by 
\begin{equation}
R_{max} \cong 3 \cdot 10^{15} M_{33}^{1/2} \zeta_{UV}^{-1/4} \, \textrm{cm} \, .
\end{equation}

In the central regions, the expansion of the gas envelopes lead to interactions between envelopes originating from different events. Further shocks are then expected, with the expanding regions filling a volume $\sim R_{max}^{3}$. 

In Paper I, we distinguished two classes of objects, Class Q and Class S, according to the importance of the volume filling factor of the configuration relative to the typical system size, defined as
\begin{equation}
\epsilon = \left( \frac{R_{max}}{100\zeta_{UV} R_{S}} \right)^{3} \cong 10^{-3} \, M_{33}^{3/2} \zeta_{UV}^{-15/4} \left( \frac{M_{BH}}{10^{9} M_{\odot}} \right)^{-3} \, .
\end{equation}

Class Q objects are characterized by a small filling factor ($\epsilon \ll 1$), a condition met for large black hole mass; Class S objects are characterized by a large filling factor ($\epsilon \sim 1$) associated with a small central mass. We identified Class Q objects with massive and luminous quasars, while Class S objects were identified with less luminous sources, such as Seyfert galaxies.
In the following we focus on Class S objects, since timing analysis are mainly performed on local Seyfert 1 galaxies. 
Each class was further subdivided into two cases, Case A and Case B, depending on whether the radiation time is longer (Case A) or shorter (Case B) than the accretion time.

The second generation shocks occur in optically thin conditions. 
In these optically thin shocks, Coulomb collisions with ions, carrying most of the gravitational energy, energetize the electrons which cool through Compton radiation. 
The seed photons for Compton upscattering are provided by the optical/UV photons emitted in the optically thick event.  
In a stationary situation, the plasma temperature is set by the balance between heating and cooling rates. 
The electron temperature is then estimated assuming an equilibrium between Coulomb heating and Compton cooling: $L_{Coulomb} = L_{Compton}$. This gives a value for the average electron energy $E_{e}$ of the order of a few hundred keV, hence Compton cooling of the hot electrons is responsible for the observed X-ray emission.
The resulting X-ray luminosities, $L_{X}$, in the different classes and sub-cases have been calculated in Paper I.


\section{X-ray variability properties}

\subsection{Characteristic time scale}

We introduce the heating time scale of the electrons given by
\begin{equation}
\tau_{heat} = \frac{E_{e}}{(\frac{dE}{dt})_{heat}} \, . 
\end{equation} 

At equilibrium this is equal to the electron Compton cooling time 
\begin{equation}
\tau_{cool} = \frac{E_{e}}{(\frac{dE}{dt})_{cool}} \, .
\end{equation} 

The above time scales are directly linked to the physical process responsible for the X-ray emission and can therefore be considered as a characteristic X-ray time scale, defined as:
\begin{equation}
\tau_{X} \sim \tau_{heat} \sim \tau_{cool} \, .
\end{equation}

The numerical value is of the order of $\sim$days:
\begin{equation}
\tau_{X} \cong 5 \, \eta_{1/3}^{-1} \zeta_{UV}^{3} \dot{M}_{0}^{-1} M_{8}^{2} \; \textrm{d} \, ,
\label{tau_value}
\end{equation}
where the black hole mass is expressed in units of $M_{BH} = M_{8} \cdot 10^{8} M_{\odot}$, and the accretion rate in units of $\dot{M} = \dot{M}_{0} \cdot 1M_{\odot}/yr$. 

The characteristic X-ray time scale is proportional to the square of the central mass and inversely proportional to the accretion rate. 
The shortest time scale is obtained in the case of a low mass object accreting at a high rate. We note that the dependence on mass is stronger than that on accretion rate. 
The dependence on black hole mass and accretion rate comes from the photon energy density given by the luminosity of the optically thick shocks, proportional to $\dot{M}$, and the average distance of the optically thick shocks at the origin of the UV photons, proportional to $M_{BH}^{2}$.
Thus X-ray variability seems to be determined by the source parameters, black hole mass and accretion rate. This may account for the different variability properties observed in different AGN classes.

\subsection{Luminosity variations}

In our picture the X-ray emission is produced by Comptonization of seed UV photons, emitted in the optically thick shocks, by energetic electrons created in an optically thin event.
The X-ray luminosity then depends on the properties of both optically thick and optically thin shocks through the photon energy density ($u_{ph}$) and the electron number density ($n_{e}$), respectively.
We therefore re-express the X-ray luminosities leaving the explicit dependences on $u_{ph}$ and $n_{e}$, using the photon energy density and the electron number density estimated in Paper I (eq. (13) and eq. (14), respectively). 

Here we parametrize these two quantities by appropriate values ($u_{ph} = u_{\gamma} \cdot 35.4 \, \mathrm{erg/cm^{3}}$ and $n_{e} = n_{8} \cdot 1.2 \cdot 10^{8} \, \mathrm{cm^{-3}}$), to obtain:
\begin{equation}
\textrm{Case A:} \quad
\left\langle L_{X} \right\rangle \cong 4.9 \cdot 10^{43} \, E_{p,MeV}^{4/7} \zeta_{UV}^{3} M_{8}^{3} u_{\gamma}^{5/7} n_{8}^{9/7} \; \textrm{erg/s} 
\end{equation} 
\begin{equation}
\textrm{Case B:} \quad
\left\langle L_{X} \right\rangle \cong 2.1 \cdot 10^{43} \, E_{p,MeV}^{4/7} \zeta_{UV}^{3/2} M_{8}^{2} u_{\gamma}^{-2/7} n_{8}^{9/7} \; \textrm{erg/s}
\label{ratio_IB}
\end{equation}

We observe that the dependence of the X-ray luminosity on the electron number density is stronger than that on the photon energy density. This already suggests that the resulting X-ray luminosity and its variations are more sensitive to variations of the upscattering medium than those in the seed photon population. 

We can now relate variations in the X-ray luminosity to electron density modulations. Assuming that the main dependence is on the electron density, we suppose a relation of the form $L_{X}(n_{e}) = K \cdot n_{e}^{9/7}$ (where $K$ is a numerical factor). We then consider small density perturbations $\Delta n_{e}$ around a mean value $n_{e}$, and expanding to first order we obtain: 
\begin{equation}
\frac{\Delta L_{X}}{L_{X}} \sim \frac{9}{7} \frac{\Delta n_{e}}{n_{e}} \, .
\label{delta}
\end{equation} 

Expression ($\ref{delta}$) directly relates density fluctuations to luminosity variations.
Modulations in the electron density can lead to significant X-ray variations of order unity, with density fluctuations being slightly amplified. 
Variations in the seed photon population would have a weaker effect. 

Density modulations typically occur on a time scale associated with the inhomogeneous structure resulting from the succession of shocks. We introduce the inhomogeneity time scale $\tau_{inh} \sim l/V$, where $l$ is a typical inhomogeneity size and $V$ the medium velocity.  
This time scale is essentially determined by the local inhomogeneity properties and does not scale with the system parameters, such as the black hole mass. 
It can cover a broad range of time scales reflecting the inhomogeneities of the accretion flow.
The slope of the luminosity variations as a function of the temporal frequency, $\frac{\Delta L_{X}}{L_{X}}(\nu=V/l)$, is related to that of the density modulations as a function of the inhomogeneity size, $\frac{\Delta n_{e}}{n_{e}}(l)$. Thus the shape of the power spectrum is given by the spectrum of density fluctuations. 
Electron density modulations can be transmitted, and thus observed as luminosity variations, only if the local inhomogeneity time is greater than the electron heating time ($\tau_{inh} > \tau_{heat}$). Fluctuations on time scales shorter than the heating time ($\tau_{inh} < \tau_{heat}$) cannot be propagated, they are smoothed out and effectively suppressed.
This leads to a decline in variability power, which induces a break in the power spectrum.
This is schematically illustrated in Figure 1.

   \begin{figure}[h!bpt]
   \centering
   \includegraphics[width=0.5\textwidth]{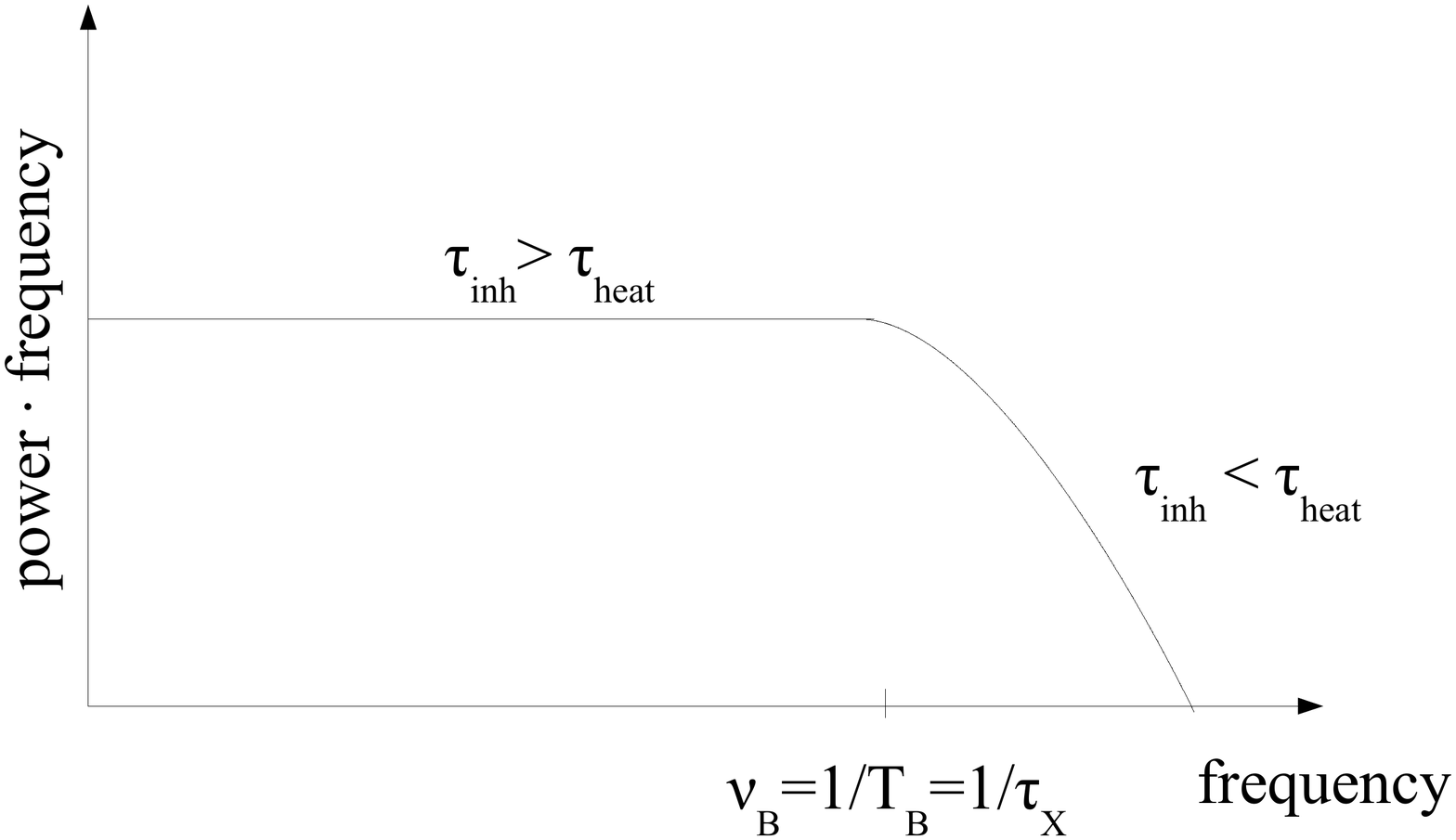}
      \caption{Sketch of a typical power spectrum in the $(\nu \cdot P_{\nu})$ vs. $\nu$ representation.
      Electron density modulations on time scales longer than the heating time ($\tau_{inh} > \tau_{heat}$) are effectively transmitted, leading to fluctuations in the emitted radiation.
This corresponds to the flat portion in the observed power spectrum.
Rapid modulations on time scales shorter than the heating time ($\tau_{inh} < \tau_{heat}$) are smoothed out, with the amplitude of fluctuations being considerably dampened. This corresponds to the steepening observed in the power spectrum.
The suppression of fluctuations on time scales shorter than a critical value leads to a break in the variability power which may be associated with the observed PSD break. 
 }      
         \label{figure}
   \end{figure}


\section{Break time scale: comparison with observations}

\subsection{Magnitude and functional dependence}

The Power Spectral Density (PSD) is defined as the modulus-squared of the Fourier transform of the light curve, in units of light curve variance per Hz. The typical power spectrum is characterized by a slope of $\alpha$$\sim$$-1$, steepening to a slope of $\alpha$$\sim$$-2$ above a break frequency $\nu_{B}$. 
A break time scale $T_{B} = 1/\nu_{B}$ is associated with the bend in the power spectrum and can be considered as a characteristic X-ray variability time scale. 

Break time scales in AGNs have been studied by many authors and PSD measurements have been obtained for around 20 sources (\citet{U_2007}).
Here we summarize a number of recent observational results. 
\citet{M_et_2003} analysed a sample of six Seyfert 1 galaxies and found a significant correlation between break time scale and black hole mass. The measured PSD break time scales were typically of the order of a few days, and the data could be fitted by a linear relation of the form $T_{B} (\textrm{days}) \approx M_{BH}/10^{6.5} M_{\odot}$. The luminosity-time scale correlation was found to be negligible compared with the mass-time scale relation. 
A similar result was obtained in the case of broad-line galaxies with the break frequency decreasing with increasing mass, following  a relation of the form $\nu_{hfb} \approx 1.5 \cdot 10^{-6}/(M/10^{7}M_{\odot})$Hz (\citet{P_2004}).
More recently, \citet{M_et_2008} reported a break time scale of $\sim$34 days in Mrk 509, in agreement with the mass-break time scale relationship discussed in the previous works. 

However, a significant scatter is observed in the linear black hole mass-break time scale correlation. 
Broad-line AGNs are consistent with a linear scaling of break time scales with central mass, but narrow-line galaxies are observed to lie systematically above this relation. 
This implies that, for a given black hole mass, the break time scale should be shorter in narrow-line galaxies than in broad-line objects. 
Following this supposition, \citet{McH_et_2004} suggested that the break time scale might depend on additional parameters such as accretion rate and/or black hole spin.
The difference in break time scales observed in different AGN classes has been confirmed by \citet{U_McH_2005} who analysed the mass-time scale relation for all available objects with measured PSDs. They showed that for a given black hole mass, higher accretion rate objects indeed have shorter break time scales than lower accretion rate counterparts. 

The main result, on which all observational works agree, is the existence of a scaling between characteristic time scale and black hole mass. The typical value of the break time scale is of the order of $\sim$days for a $10^{7}-10^{8} M_{\odot}$ object.
At fixed central mass, the break time scale is shorter in higher accretion rate systems, such as NLS 1 galaxies. 
The numerical value of the X-ray time scale defined in eq. ($\ref{tau_value}$) thus matches the typically measured $T_{B}$ value. This correspondence leads us to associate the characteristic X-ray time scale $\tau_{X}$ with the PSD break time scale $T_{B}$.   
The predicted scaling of $\tau_{X}$ with black hole mass is in agreement with the observed mass-break time scale correlation. For a given black hole mass $\tau_{X}$ decreases with increasing accretion rate, a trend confirmed by the short time scales observed in narrow-line objects. 

The general relation between break time scale, black hole mass, and accretion rate has been 
quantified by \citet{McH_et_2006}. Following the hypothesis that the break time scale depends on both black hole mass and accretion rate, they obtained a scaling relationship of the form
\begin{equation}
T_{B} \approx \frac{M_{BH}^{1.12}}{\dot{m}_{E}^{0.98}} \, ,
\label{eq_McH}
\end{equation}
where $\dot{m}_{E} \approx L_{bol}/L_{E}$, with $L_{bol}$ the bolometric luminosity and $L_{E}$ the Eddington luminosity. 
The Eddington luminosity scales with black hole mass ($L_{E} \propto M_{BH}$) and assuming that the bolometric luminosity is proportional to the accretion rate ($L_{bol} \propto \dot{M}$), the empirical relationship can be re-written as:
\begin{equation}
T_{B} \propto \frac{M_{BH}^{2.1}}{\dot{M}^{0.98}} \, .
\label{T_B} 
\end{equation} 

We recall that the predicted dependence of the characteristic X-ray time scale on central mass and accretion rate is of the form:
\begin{equation}
\tau_{X} \propto \frac{M_{BH}^{2}}{\dot{M}} \, .
\label{tau_dependence}
\end{equation} 

The model dependence is therefore in excellent agreement with the observational relation obtained by \citet{McH_et_2006}.  \\

The empirical relation given in eq. ($\ref{eq_McH}$) is stated to be valid down to galactic black hole binary systems in which accretion is expected to be ruled by angular momentum dissipation in a disc. 
We will investigate whether our model results can be extended to this regime in a future work.

\subsection{Physical interpretations}

Several authors have proposed possible interpretations of the X-ray characteristic time scale by comparing different model time scales with the observed break time scales.
In this perspective, different accretion disc time scales have been compared with measured PSD break time scales (\citet{M_et_2003}, \citet{P_2004}). The orbital, thermal, and viscous time scales at different radial distances have been considered. Orbital time scales are too short; on the contrary, viscous time scales are much too long. The thermal time scale, at certain radii, gives the closest value to the measured break time scale. In such cases, thermal instabilities have been invoked as a possible source of variability. 

The characteristic variability time scale has also been discussed in the framework of inner propagating accretion flow fluctuation models (\citet{L_1997}, \citet{C_et_2001}). 
In this picture, modulations produced at different outer radii in the accretion flow propagate inwards, until reaching the inner X-ray emitting region where they cause variations of the X-ray flux. These authors consider a geometry formed by a geometrically thin, optically thick disc surrounded by a geometrically thick corona extending out to large radial distances. The disc is assumed to be truncated at some radius from the centre. 
The characteristic X-ray time scale is then associated with the time scale of the modulations at the truncation radius of the disc. The difference in break time scale can be related to the location of the disc truncation radius: the smaller the truncation radius the shorter the characteristic time scale.

In our model, the PSD break time scale is associated with the characteristic X-ray time scale, $\tau_{X}$. Since $\tau_{X}$ represents the heating/cooling time of electrons, the break time scale is directly linked to the physical process at the origin of the observed X-ray emission. Its numerical value and functional dependence are in agreement with the observed PSD break time scale. 
Fluctuations on time scales shorter than the heating time cannot be transmitted and the resulting suppression of rapid variations translates into a break in the power spectrum. The observed PSD break then reflects the suppression of short time scale variations.
We are thus able to predict, at least qualitatively, the overall shape of the power spectrum and in particular the appearance of the break.

Below the break frequency, variations are found to hold the same slope ($P_{\nu} \propto \nu^{-1}$, flicker noise) over several decades in frequency. For instance, the slope in NGC 4051 remains unchanged for over four decades (\citet{McH_et_2004}).
The similar shape suggests that the same physical mechanism is responsible for the variations over this broad range of frequencies or equivalently time scales. 
In the inner propagating fluctuations model, this broad range of variability time scales is provided by the range of radii where modulations initially originate. Since different radial distances are associated with different time scales, the optically thin corona must be extended in the radial direction in order to account for the observed range of time scales.
In the picture discussed here, the broad range of time scales can be provided by the range of time scales associated with local inhomogeneities in the accretion flow. 
The inhomogeneity time is independent of the system parameters, while the heating time is determined by the properties of the source. In particular the latter is shorter for small central mass and/or high accretion rate objects. Since the inhomogeneity time does not scale with the source parameters, the condition $\tau_{inh} > \tau_{heat}$ is easily met in such objects. Therefore the decline in variability (observed as a break in the power spectrum) occurs on a shorter time scale, explaining the large rapid variability characteristic of NLS 1 galaxies.


\section{Optical/UV variability}
 
X-ray variability is also related to variations in the seed photon population. It is thus important to investigate possible sources of seed variations and to study the related optical/UV variability.  
In general, optical/UV light curves are characterized by irregular and aperiodic fluctuations. 
This suggests that variability is due to random processes and stochastic approaches seem appropriate.  
In the framework of the so-called discrete event models, variability is attributed to a superposition of independent and random events described by a Poisson distribution.  
The physical nature of the events can be represented by a variety of phenomena: supernova explosions, stellar collisions, and hot spots on the accretion disc surface.
A characteristic variability time scale is given by the event duration, part of the observed UV variability is then directly related to the event rate. 

Characteristic variability time scales may be obtained from the study of time series through power spectral densities and structure function analysis. 
\citet{C_P_2001}  derived a characteristic optical/UV time scale of $\sim$5-100 days in a sample of 10 AGNs, performing a structure function analysis.
Studying the variability time scales in the framework of discrete-event models, \citet{F_et_2005} showed a lack of (strong) dependence of the event duration on the source luminosity, hence on central mass. In their sample, the average luminosity of the sources covered four orders of magnitude, while the event duration varied by only two orders of magnitude. Physical interpretations based on accretion disc time scales, which scale linearly with black hole mass, can thus be excluded. 
In our model, the typical time scale for the radiation of the shock energy is given by the expansion time. A characteristic UV time scale can then be associated with this expansion time, which is of the order of $\tau_{exp} \cong 12 M_{33}^{1/2} \zeta_{UV}^{1/4} \, \textrm{d}$.
The order of magnitude lies in the observed range and is determined by collision parameters. It has no explicit dependence on the black hole mass, in agreement with observations of \citet{F_et_2005}.  Shocks can be described by a Poisson distribution with each event rising and decaying on its own characteristic time scale given by the expansion time. These long-term variations may be associated with large amplitude variations seen in the optical/UV range over hundreds of days. 

Discrete event models, in which a more luminous object is obtained simply by increasing the number of identical events, predicts a power law slope of -1/2 for the variability versus luminosity relation ($\sigma(L) \propto L^{-1/2}$).
But this value is found to be incompatible with the observed trend which seems to favor a flatter dependence, with a slope of -0.08 (\citet{P_C_1997}). 
Two possibilities are envisaged: either the variability process does not follow a Poisson distribution or the individual events are not identical, with the latter hypothesis being more plausible. 
Indeed, \citet{F_et_2005} found a lack of correlation between event rate and object luminosity, suggesting that a larger total luminosity cannot only  be explained by a greater number of identical events. This implies that the event energy cannot be the same for all events. \\
In our picture, the average UV luminosity is given by the luminosity of a single event multiplied by the average number of collisions: $\left\langle L_{UV} \right\rangle = L_{UV} \cdot \left\langle N_{c} \right\rangle$. 
If the difference in total luminosity is not due to the difference in the number of events, then single events should produce different luminosities. 
In Paper I, we have derived the UV luminosity of a single event in terms of the collision parameters:
\begin{equation}
L_{UV} \cong 3 \cdot 10^{45} \, \eta_{1/3} M_{33}^{1/2} \zeta_{UV}^{-5/4} \, \textrm{erg/s} \, .
\end{equation}

The luminosity radiated in each shock is different, as each event is characterized by different collision parameters. 
In this case the difference in total luminosity is attributed to differences in the luminosity of individual events and not to the event rate. This can be reconciled with the observed flat dependence of variability on luminosity. 

As the first shock occurs in optically thick conditions, the photospheric temperature can be estimated assuming blackbody emission. The post-shock temperature evolution has been studied by \citet{C_T_2005}. 
The predicted time delays between optical and UV light curves (increasing with the increase of the wavelength difference) correspond to the observed time lags.  
Moreover a very short lag between optical and UV is expected in our model since the optical/UV emissions share a common origin, arising from the same optically thick event. This property is in agreement with the observed quasi-simultaneity of optical and UV light curves (\citet{C_C_1991}). 
The underlying common origin also seems to be supported by the observed equivalence of the mean optical and ultraviolet PSDs (\citet{C_P_2001}). 

An additional source of variability may be related to the geometry of the expansion following the optically thick shock. We have assumed a spherical expansion for simplicity, which is a very rough approximation. In reality, the expansion is expected to occur in a more irregular and asymmetric way, with different elements of the expanding envelopes having different speeds and thus providing a range of time scales. The inhomogeneous expansion may then account for shorter time scale variations, and this could explain the short-term variability sometimes observed even in the optical/UV.

The multiple contributions to the UV variability then affect the X-ray variability, since optically thin shocks arise from optically thick ones, providing a connection between the two energy bands.

\section{Correlations between optical/UV and X-ray emissions}

The study of correlations between different energy bands are thought to provide information concerning emission processes and causal links relating different emission regions. 
Indeed, the UV and X-ray emissions are coupled through reprocessing of X-rays and/or upscattering of UV seed photons. 
Time lags with longer wavelengths leading shorter ones have been interpreted as fluctuations in the accretion disc, propagating from the outer optical/UV emitting regions toward the innermost X-ray emitting region, on the viscous or thermal time scale.
Several multi-wavelength monitoring campaigns have searched for correlations and time lags between the two energy bands, in a number of AGNs.
However, the sign of the time lag can be different from case to case, and even the existence of correlations is not always confirmed (\citet{N_et_1998}, \citet{M_et_2002}, \citet{M_et_2008}). The global picture is still quite confusing and the problem seems not definitively settled yet. 

In our cascade model, the two emitting media are coupled through the succession of shocks, which also determines the sign of the time lag : optically thick shocks provide the seed photons that will later be upscattered by electrons heated in the optically thin shocks. 
Following the optically thick shock, UV photons are emitted and escape the region when the expanding clump reaches the size of the photosphere and becomes optically thin, i.e. after $\tau_{exp}$ from the shock event. 
The expansion continues until neighboring envelopes overlap, leading to optically thin shocks in which electrons gain energy on the characteristic heating time. 
X-rays are emitted as a result of the interaction between photons, generated in the current event, and hot electrons, created in a previous optically thin event. 
Two distinct time lags are then expected: a nearly zero lag due to the immediate Comptonization process, and a longer lag related to the temporal evolution of electrons, with the UV photons leading the X-rays.  
The first lag between UV and X-rays is very small, being of the order of the light travel time. 
In Class S, the second lag is expected to be of the order of the heating time, since optically thin shocks occur rapidly after the expanding sphere has become optically thin.
A rough estimate of the order of magnitude of the time lag between the two energy bands is given by $\tau_{X} \sim 5 M_{8}^{2}/\dot{M}_{0}$ d, of the order of a few days for a $10^{8} M_{\odot}$ object. 
A shorter lag is expected in sources with a smaller black hole.

Significant correlations between X-rays and optical light curves with a delay consistent with zero lag have been recently reported in MR 2251-178 (\citet{A_et_2008}) and Mrk 79 (\citet{B_et_2009}). Such correlations at very short time lags between X-ray and optical emissions may be interpreted as a result of the Comptonization process of seed photons.

Several cases with time lags of the order of $\sim$days with the optical/UV leading the X-rays  have also been observed (\citet{S_et_2003}, \citet{A_et_2005}, \citet{M_et_2008}).
The direction of the time lag and the order of magnitude are thus compatible with our predicted lags.  
\citet{M_et_2008} report a time lag of 15 days in Mrk 509, with the optical leading the X-rays. They attribute the shorter lags found in other objects (of 1-2-days for less massive Seyferts) to a lower black hole mass.
This is consistent with the above discussion, if the measured lags are associated with lags related to the heating time. 

In Class Q, where the filling factor is small, we need to take into account the different locations of optically thick and optically thin shocks, with the corresponding electron travel time.
Matter has to travel from the 100 $R_{S}$ region where seed photons are emitted toward the innermost region where optically thin shocks take place. 
This is done on the local free-fall time, which is of the order of several months for a $10^{9} M_{\odot}$ source. 
We would then expect an order of magnitude longer time lags in massive objects compared with Class S sources.
Cross-correlation analysis of 3C 273 show two peaks between UV and X-ray light curves, one at zero lag and the other with the UV leading the X-rays by $\sim$1.8 yr (\citet{P_C_W_1998}, \citet{S_et_2008}). 
This suggests that the first lag corresponds to the Comptonization process, while the second one may be interpreted as the travel time, from the 100$R_{S}$ region to the innermost region. 
Thus the overall picture seems to be supported by observations. 
However, one has to bear in mind that other physical processes might contribute to the observed properties (such as particular features possibly related to a jet component) leading to non negligible uncertainties.


\section{Conclusion}

We have considered characteristic time scales derived in the framework of clumpy accretion flows. The value of the characteristic X-ray time scale  $\tau_{X}$, associated with the electron heating/cooling process, corresponds to the typically observed value of the PSD break time scale, $T_{B}$. The predicted time scale is shorter for small black hole mass and/or high accretion rate, correctly reproducing the observed trend. 
The dependence on black hole mass and accretion rate we derived in eq. (\ref{tau_value}) remarkably agrees with the observational relation found by \citet{McH_et_2006}, without additional parameters. 

In our picture, X-ray variability is attributed to both variations in the seed photon population and density fluctuations in the upscattering medium, with a greater contribution from the latter.
A break is expected in the power spectrum, since fluctuations on time scales shorter than the electron heating time are not observable. The associated X-ray time scale $\tau_{X}$, directly related to the physical process of X-ray emission, may thus provide a possible interpretation of the PSD break time scale.



\bibliographystyle{aa}
\bibliography{biblio}

\end{document}